\documentclass[12pt,thmsa]{article}
\usepackage{amssymb}
\usepackage{graphicx}

 \makeatletter
\newcommand{\row}[1]%
{\mathord{\buildrel{\lower3pt%
\hbox{$\scriptscriptstyle\rightarrow$}}\over #1}}

\newcommand{\dyadic}[1]{\mathord{\dyadic@rrow{#1}}}
\newcommand{\dyadic@rrow}[1]{
\begin{picture}(12,12)(-1,0)
\put(-2,10){\makebox(0,0)[t]{$\scriptscriptstyle\downarrow$}}
\put(-2,11){\makebox(0,0)[l]{$\scriptscriptstyle\longrightarrow$}}
\put(5,0){\makebox(0,0)[b]{$#1$}}
\end{picture}
}

\newcommand{\bra}[1]{\bigl\langle #1 \bigr|}
\newcommand{\ket}[1]{\bigl| #1 \bigr\rangle}

\begin{document}
\begin{center}
\textbf{ Estimation of  pulsed driven qubit parameters via quantum
Fisher information }

 N. Metwally and S. S. Hassan\\
 Department of Mathematics, College of Science,
 University of Bahrain, P. O. Box  32038 Kingdom of  Bahrain\\
 {\small Nmetwally@gmail.com, Shoukryhassan@hotmail.com}

\end{center}
\begin{abstract}
We estimate the initial weight and phase  parameters ($\theta,
\phi)$ of a single qubit system initially prepared in the coherent
state $\ket{\theta,\phi}$ and  interacts with three different
shape of pulses; rectangular, exponential, and $sin^2$-pulses. In
general, we show that the estimation degree of the weight
parameter depends on the pulse shape and the initial phase angle,
$(\phi)$. For  the rectangular pulse case, increasing the
estimating rate of the weight parameter via the Fisher information
function $(\mathcal{F}_\theta)$ is possible with small values of
the atomic detuning parameter and larger values of the pulse
strength.
 Fisher information $(\mathcal{F}_\phi)$ increases suddenly at
resonant case to reach its maximum value if the initial phase
$\phi=\pi/2$ and consequently one may estimate the phase parameter
with high degree of precision. If the initial system is coded with
classical information, the upper bounds of Fisher information for
resonant and non-resonant cases are  much larger and consequently
one may estimate the pahse parameter with high degree of
estimation. Similarly as the detuning increases the Fisher
information decreases and therefore the possibility of estimating
the phase parameter decreases. For exponential, and $sin^2$-pulses
the Fisher information  is maximum
($\mathcal{F}_{\theta,\phi}=1$) and consequently one can always
estimate the weight and the phase parameters $(\theta,\phi)$ with
high degree of precision.

\end{abstract}
\section{Introduction}

The essence of conditional probability is that, learning about one
event (or measurement) changes or affects the probability for a
second event. Classically, this is well known termed as Fisher
information (FI) function \cite{Barnett}. In the other words, the
FI extracts information about an unknown parameter, $\beta$, from
a previously measured result result, say, $x$.

For quantum systems, the corresponding quantum Fisher information
(QFI) is nowadays an important physical quantity of estimation
within the context of quantum metrology and quantum information
theory \cite{Maccone}. It describes the sensitivity of a quantum
state with respect to changes in its initial parameters
\cite{Jian011,Yao014} or gained system parameters during the
quantum information processing as teleportation
\cite{Metwally017}. Evaluation of QFI with respect to the desired
estimation parameters have been given in \cite{Wei013,Jing013}.
Dynamics of QFI were studied in many models, such as, the Ising
model\cite{Wang011}, mixed Hamiltonian model \cite{Hu014}, Bell
state pairs model under decoherence channel \cite{Ozaydin014} and
steady state   open and noisy systems \cite{Ali016}. QFI in
non-inertial frames has been also  quantified for different
systems. For example, Yao et. al \cite{yao014-1} have investigated
the dynamics of Fisher information of a pure-two qubit state in
non-inertial frame. Metwally \cite{Metwally017} has investigated
the dynamics of the teleported QFI by using accelerated quantum
channel. The Unruh acceleration effect on the precision of
parameter estimation for a general two qubit system is discussed
in\cite{Metwally017-1}.

Investigating QFI  of a driven  single qubit by a laser pulse is
important  within the  context of quantum information theory.
Previously, we have  investigated the transfer and exchange
information between  a single qubit system  and the driving
rectangular pulse \cite{Metwally012}. Also, the possibility of
achieving  long-lived entanglement between two entangled pulsed
driven qubits  is given in \cite{Metwally014}.

Here, we extend our investigation to the behavior of QFI of a
single qubit driven by different shapes of pulses. Specifically,
we investigate the problem of parameter estimation of the quantum
channel parameter for a single qubit, initially prepared in a
coherent state and driven by three different shapes of pulses ,
namely, rectangular, exponential and $\sin^2$ pulses.

The paper is organized as follows. In Sec.2, we present the model
and its  exact  solution   in terms of its Bloch vectors, together
with the definition of the QFI. Computational results of the
effect of the pulse shapes on the QFI are represented in Sec.3,
followed by a summary in Sec.4.

\section{The suggested Model}
\subsection{The Hamiltonian Model}
Her, we consider a single qubit taken as 2-level atomic transition
of frequency $\omega_q$ and driven by a laser pulse of arbitrary
shape and of circular frequency $\omega_c$ in the absence of any
dissipation process. The quantized Hamilationan of the system (in
units of $\bar h=1$) in the dipole and rotating wave approximation
is given by\cite{Sukry008, Hassan010}

\begin{equation}\label{Ham}
\hat{H}=\omega_q\hat{S_z}+\frac{\Omega(t)}{2}(\hat{S_{+}}e^{-i\omega_c
t}+\hat{S_{-}}e^{i\omega_c t})
\end{equation}
where,  the spin-$\frac{1}{2}$ operators $\hat{S_{\pm},z}$  obey
the $Su(2)$ algebra,
\begin{equation}\label{Com}
[\hat{S_{+}}, \hat{S_{-}}]=2\hat{S_{z}},\quad
[\hat{S_{z}},\hat{S_{\pm}}]=\pm\hat{S_{\pm}}
\end{equation}
and $\Omega(t)=\Omega_o f(t)$,is the real laser Rabi frequency
with $f(t)$ is the pulse shape. Introducing the rotating frame
operators,

\begin{equation}
\hat{\sigma_{\pm}}(t)=\hat{S_{\pm}}(t)e^{\mp\omega_c t}, ~\quad
\hat{\sigma_z}(t)=\hat{S_z}(t)
\end{equation}
where  the $\hat{\sigma}$ operators  obey the same algebraic form
of Eq.(\ref{Com}. Heisenberg equations of motion for the atomic
operators $\hat{\sigma}_{\pm,z}$  according to (\ref{Ham}) are of
the form,
\begin{eqnarray}\label{motion}
\dot{\hat{\sigma}}_{+}&=
&i\Delta\hat{\sigma}_{+}-i\Omega(t)\hat{\sigma}_z
=\Big[\dot{\hat\sigma}_{-}\Bigl]^\dag,
\nonumber\\
\dot{\hat\sigma}_{z}&=&-i\frac{\Omega(t)}{2}
(\hat{\sigma}_{+}-\hat{\sigma_{-}})
\end{eqnarray}
where $\Delta= \omega_q-\omega_c$, is the  atomic  detuning.
Eqs(4) are  of variable coefficient and have exact solutions in
the following two cases \cite{Sukry008, Hassan010}:
\begin{enumerate}
\item Arbitrary atomic detuning $(\Delta)$ and constant Rabi
frequency $\Omega(t)=\Omega_0$ with $f(t)=1$. This case
corresponds to the  rectangular pulse shape.

\item Exact atomic resonance ($\Delta=0)$ and arbitrary Rabi
frequency $\Omega(t)$. This case corresponds to a pulse of
arbitrary shape, $f(t)$.

\end{enumerate}
Initially, we assume the qubit is prepared in the coherent state,

\begin{equation}
\ket{\psi_q}=\cos(\theta/2)\ket{0}+e^{-i\phi}\sin(\theta/2)\ket{1},
\end{equation}
where $0\leq \phi\leq 2\pi$, $0\leq \theta \leq \pi$ and
$\ket{0},\ket{1}$ are the lower and  upper states, respectively.
We use  the notations,

\begin{equation}
s_i(0)=\bra{\theta,\phi}\hat{\sigma_i}\ket{\theta,\phi}
\end{equation}
for  the initial  Bloch vector components, where   ($i=x,y,z$) and
$\hat\sigma_{x}=\frac{1}{2}(\hat{\sigma_{+}}+\hat{\sigma_{-}}),~
\hat\sigma_y=\frac{1}{2i}(\hat\sigma_{+}-\hat{\sigma_{-}})$.
Explicity from (5) and (6), we have,
\begin{equation}
s_x(0)=\sin\theta\cos\phi, ~~s_y(0)=\sin\theta\sin\phi,~~
s_z(0)=-\cos\theta
\end{equation}

\subsection{Exact Solutions}
In the light of the comments (i),(ii) after Heisenberg eq (4), we
now present their exact solutions in the following three cases of
the pulse shape:

\begin{enumerate}
\item  Rectangular pulse:\\
 For a  rectangular laser  pulse of duration $T$,  the Rabi
frequency  $\Omega(t)=\Omega_0 f(t)$, where $\Omega_0$ real and
$f(t)$ is defined as,

\begin{equation}\label{RectPulse}
 f(t) = \left\{ \begin{array}{ll}
1&  \textrm{ for  $t\in[0,T]$}\\
0 & \quad\textrm{otherwise}\\
\end{array} \right.
\label{equation6}
\end{equation}
where the pulse duration $T$ is much shorter than the lifetime of
the qubit upper state, hence the  damping  can be discarded. The
exact solutions of (4)  in terms of the Bloch vector
$s_{x,y,z}(t)$ are given  in the form \cite{Metwally012,Sukry008},

\begin{eqnarray}
s_x(t)&=&\Bigr[\Bigr(\frac{1}{\eta}+\delta^2
\cos(\tau\sqrt{\eta})\Bigl]
-\frac{\delta\sin(\tau\sqrt{\eta})}{\sqrt{\eta}}\Bigl]
 s_x(0)
 \nonumber\\
&&+\Bigr[\frac{1+2\delta^2}{2\eta}+\frac{\cos(\tau\sqrt{\eta})}{2\eta}
+\frac{\delta\sin(\tau\sqrt{\eta})}{\sqrt{\eta}}\Bigl]s_y(0)
\nonumber\\
&&+\Bigr[\frac{\delta}{\eta}\Bigr(1-\cos(\tau\sqrt{\eta})\Bigl)
+\frac{1}{\sqrt{\eta}}\sin(\tau\sqrt{\eta})\Bigl]s_z(0),
\nonumber\\
 s_y(t)&=&\Bigr[\Bigr(\frac{1}{2\eta}+
 \frac{\eta+\delta^2}{2\eta}\cos\tau\sqrt{\eta}\Bigl)
 +\frac{1}{\sqrt{\eta}}\sin(\tau\sqrt{\eta})\Bigl]s_x(0)
 \nonumber\\
&&+\Bigr[\cos(\tau\sqrt{\eta})-
 \frac{\delta}{\sqrt{\eta}}\sin(\tau\sqrt{\eta})\Bigl]s_y(0)
 \nonumber\\
 &&+\Bigr[\frac{\delta}{\eta}\bigr(1-\cos(\tau\sqrt{\eta})\Bigl)
 -\frac{\delta}{\sqrt{\eta}}\sin(\tau\sqrt{\eta})\Bigl]s_z(0),
 \nonumber\\
s_z(t)&=&\frac{\delta}{\eta}\Bigr(1-\cos(\tau\sqrt{\eta})\Bigl)s_x(0)
+ \frac{1}{\sqrt{\eta}}\sin(\tau\sqrt{\eta})s_y(0) \nonumber\\
&&+\Bigr(\frac{\delta^2}{\eta}+\frac{1}{\eta}\cos(\tau\sqrt{\eta})\Bigl)s_z(0)
\end{eqnarray}
where $\delta=\frac{\Delta}{\Omega_0},~\eta=1+\delta^2$,~
$\tau=\Omega_0 t$ is the scaled time and $s_{x,y,z}(0)$ are given
in (7).

\item Exponential pulse:\\
In this case   $\Omega(t)=\Omega_0~f(t)$, where $\Omega_0$ is the
Rabi frequency associated with the laser pulse and the pulse shape
$f(t) $ is defined as,
\begin{equation}
 f(t) = \left\{ \begin{array}{ll}
e^{-\gamma_p t}&  \textrm{ for  $t\geq 0$}\\
0 & \quad\textrm{for $t<0$}\\
\end{array} \right.
\label{equation6}
\end{equation}
where $\gamma^{-1}_P$ is the (short) time scale of the pulse. At
exact resonance ($\Delta=0)$, the exact solution for the Bloch
vector components are given by \cite{Bat012,Bat017},
\begin{eqnarray}
s_x(t)&=&s_x(0),
 \nonumber\\
s_y(t)&=&\cos\omega_{e}(t) s_y(0)-\sin\omega_{e}(t)s_z(0)
\nonumber\\
s_z(t)&=&\cos\omega_{e}(t)s_z(0)+\sin\omega_{e}(t)s_y(0)
\end{eqnarray}
where
\begin{equation}
\omega_e(t)=\frac{\Omega_0}{\gamma_p}(1-e^{-\gamma_p t})
\end{equation}

\item {$\sin^2$-pulse:\\} In this case  $\Omega(t)=\Omega_0 f(t)$
and  the laser pulse shape $f(t)$ is given by
\begin{equation}
f(t)=sin^2(n\omega_q(t)); n=1,2,...
\end{equation}
with $(n\omega_q)$ is the beating frequency of the pulse. The
pulse shape (13) represents $n-$sequential pulses, each of
duration $\pi$ over the interval $[0,~n\pi]$. At exact resonance
$(\Delta=0)$, the Bloch vector components have the same form as in
(11), but with the time-dependent frequency $\omega_e(t)$ is
replaced by $\omega_s(t)$ \cite{ al2015,al2016}, where

\begin{equation}
\omega_s(t)=\frac{\Omega'}{2}\Bigl(\tau-\frac{1}{2n}\sin(2n\tau)\Bigr),~n=1,2,3,...
\end{equation}
where $\tau=\omega_q t$ is the normalized time and
$\Omega'_0=\frac{\Omega_0}{\omega_q}$ is the normalized pulse
strength.

\end{enumerate}

\subsection{Quantum Fisher Information}
The density operator for 2-level atomic system is given by,
\begin{equation}
\rho_q=\frac{1}{2}(I+\row{s}\cdot\row{\sigma})
\end{equation}

where, $\row{s}=(s_x(0),s_y(0),s_z(0))$  is the  Bloch vector and
$\hat\sigma=(\hat\sigma_x,\hat\sigma_y,\hat\sigma_z)$ are the spin
Pauli operators. In terms of Bloch vector $\row{s}(\beta)$, the
QFI  with respect to the parameter $\beta$ is defined as
\cite{Wei013,Xing016},

\begin{equation}
\mathcal{F}_{\beta} = \left\{ \begin{array}{ll}
\frac{1}{1-\bigl|\row{s}(\beta)\bigr|^2}\Bigl[\row{s}(\beta)\cdot\frac{\partial{\row{s}(\beta)}}{\partial\beta}\Bigr]
+\Bigl(\frac{\partial\row{s}(\beta)}{\partial\beta}\Bigr)^2&  \textrm{ for mixed state},|\row{s}(\beta)|<1,\\
\nonumber\\
\Bigl|\frac{\partial\row{s}(\beta)}{\partial\beta}\Bigr|^2 & \quad\textrm{for pure state}, ~|\row{s}(\beta)|=1\\
\end{array} \right.
\end{equation}
where $\beta$ is the parameter to be estimated. From Eq.(7), it is
clear that the final solution depends on the initial parameters
($\theta, \phi$) in addition to the system parameters $\delta$,
$\Omega'_0$.
 In the
following subsections, we shall estimate these parameters by
calculating their corresponding QFI, $\mathcal{F}_\beta$. The
larger QFI is the higher degree of estimation for the parameter
$\beta$.
\section{Computational results of QFI}
\begin{figure}[b!]
  \begin{center}
  \includegraphics[width=15pc,height=12pc]{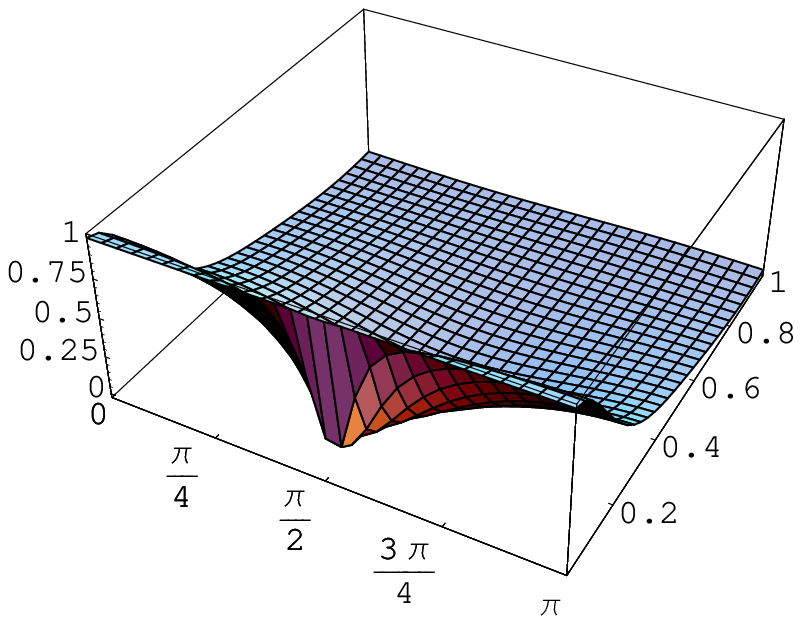}~\quad
   \includegraphics[width=15pc,height=10pc]{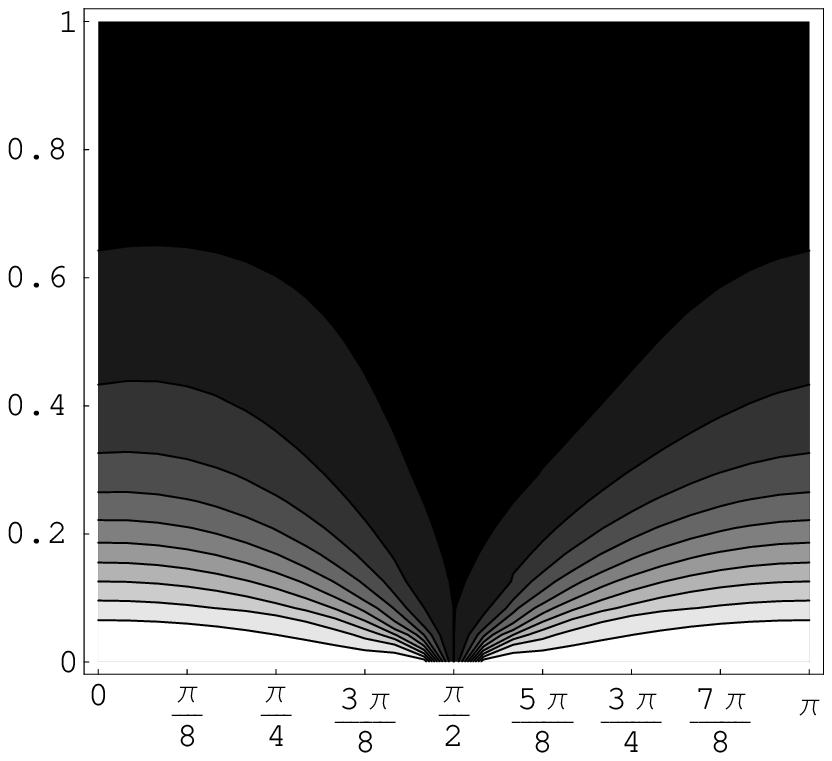}
\put(-390,68){$\mathcal{F}_\theta$}
    \put(-350,130){$(a)$}
    \put(-110,130){$(b)$}
        \put(-320,15){$\theta$}
     \put(-215,40){$\delta$}
      \put(-90,-8){$\theta$}
        \put(-180,60){$\delta$}\\
\includegraphics[width=15pc,height=12pc]{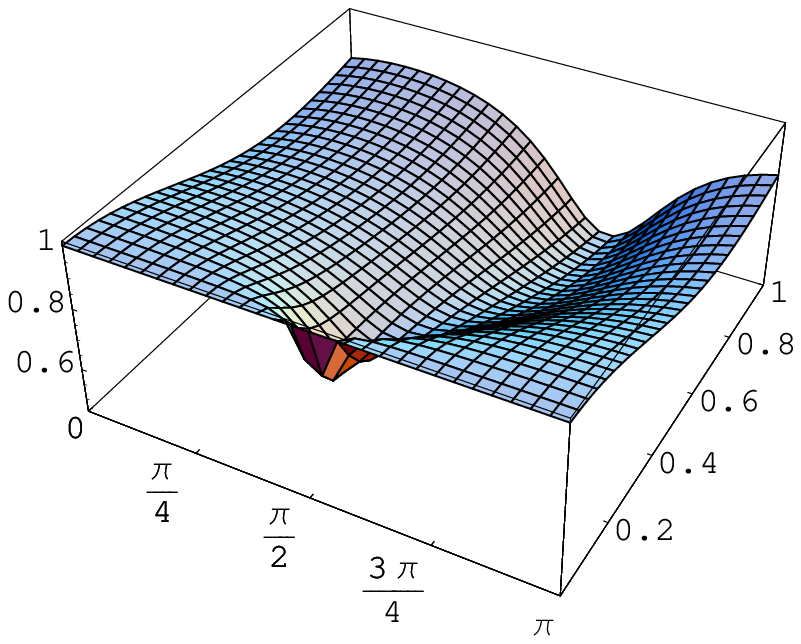}~\quad
   \includegraphics[width=15pc,height=10pc]{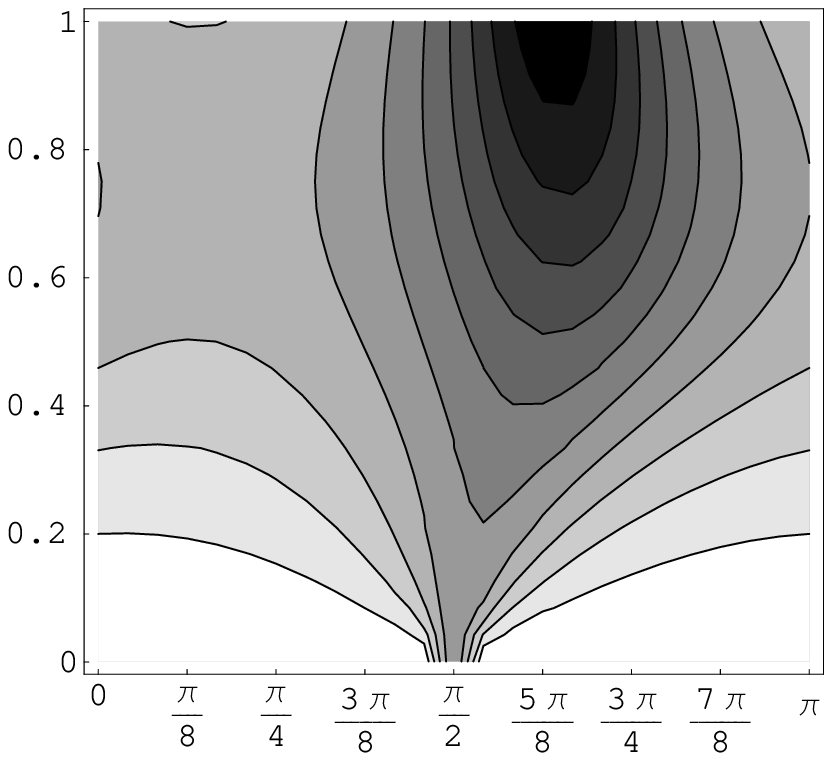}
   \put(-350,130){$(c)$}
    \put(-110,130){$(d)$}
    \put(-320,10){$\theta$}
     \put(-215,40){$\delta$}
      \put(-90,-8){$\theta$}
        \put(-180,60){$\delta$}
   \put(-390,68){$\mathcal{F}_\theta$}
       \caption{Fisher information $(\mathcal{F}_{\theta})$  in the rectangular pulse case with respect to the parameter $\theta$ and  the
   detuning parameter $\delta$ for fixed  $\phi=\pi$,  and $\Omega'_0=0.3,0.9$ for the figures (a,b),
   (c,d),   respectively.}
       \end{center}
\end{figure}
\subsection{Rectangular Pulse}
\begin{enumerate}
\item{Estimating the weight parameter $(\theta)$}

 Fig.(1) shows the behavior of
$\mathcal{F}_{\theta}$ against the detuning parameter where we
assume that the single qubit system is initially prepared in the
state $\ket{\psi_q}=\cos(\theta/2)\ket{0}-i\sin(\theta/2)\ket{1}$,
namely, we set the phase angle $\phi=\pi$. The general behavior
shows that $\mathcal{F}_\theta$ decays as the detuning parameter
increases.  Moreover, the decay is less by increasing the pulse
strength $\Omega'_0$. Fig1.(a,b)  describe the behavior of the
Fisher information  $\mathcal{F}_\theta$ at $\Omega'_0=0.3$. It is
shown that at $\delta=0$ and $\theta=0$, i.e,
$\ket{\psi_q}=\ket{0}$,  $\mathcal{F}_\theta$ is maximum. However
as one increases $\theta$ at zero detuning $(\delta=0)$ the Fisher
information $\mathcal{F}_\theta$ decays gradually to reach its
minimum value  at $\theta=\pi/2$. In this case the initial single
qubit system reduces to be
$\ket{\psi_q}=\frac{1}{\sqrt{2}}(\ket{0}-i\ket{1})$. For further
values of $\theta\in[\pi/2,~\pi]$, $\mathcal{F}_\theta$ increases
gradually to reach its maximum value at $\theta=\pi$. This maximum
value is reached for a qubit system  initially prepared in
$\ket{\psi_q}=-i\ket{1}$. However, the Fisher information
$\mathcal{F}_\theta$ decays gradually as $\delta$ increases and
starts to disappear for $\delta>0.6$.

\begin{figure}[t!]
  \begin{center}
 \includegraphics[width=15pc,height=12pc]{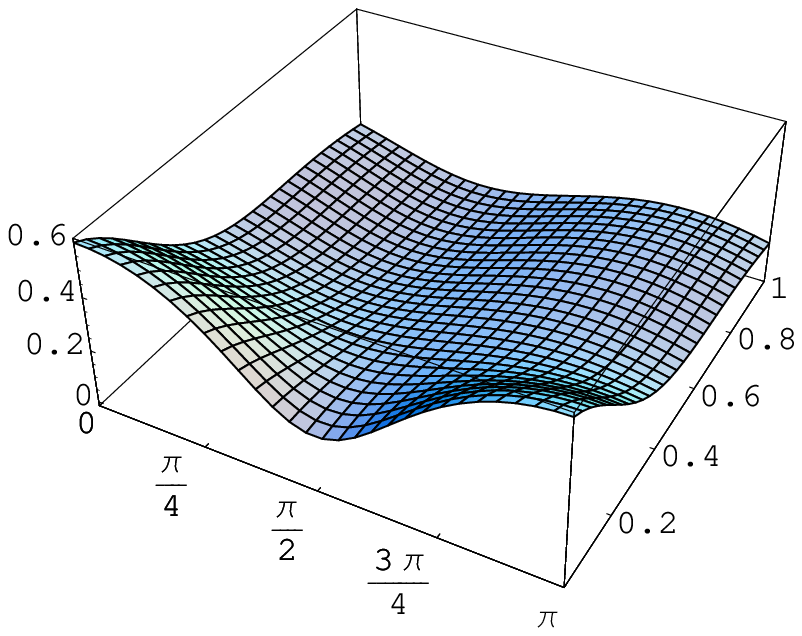}~\quad
   \includegraphics[width=15pc,height=10pc]{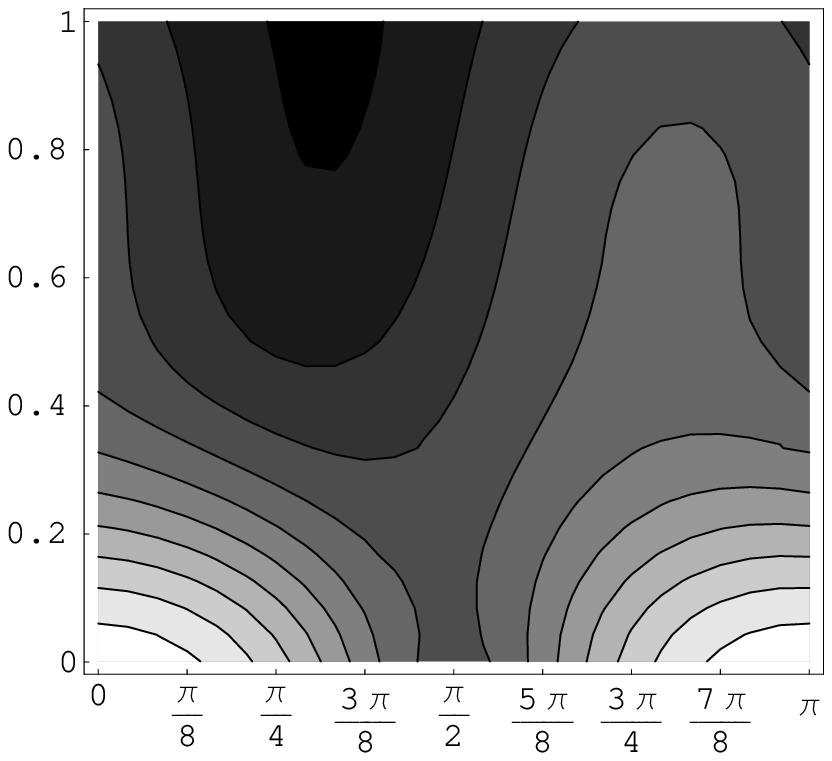}
\put(-385,68){$\mathcal{F}_\theta$}
    \put(-350,130){$(a)$}
    \put(-110,130){$(b)$}
        \put(-320,15){$\theta$}
     \put(-210,40){$\delta$}
      \put(-90,-9){$\theta$}
        \put(-180,60){$\delta$}~\\
        \includegraphics[width=15pc,height=12pc]{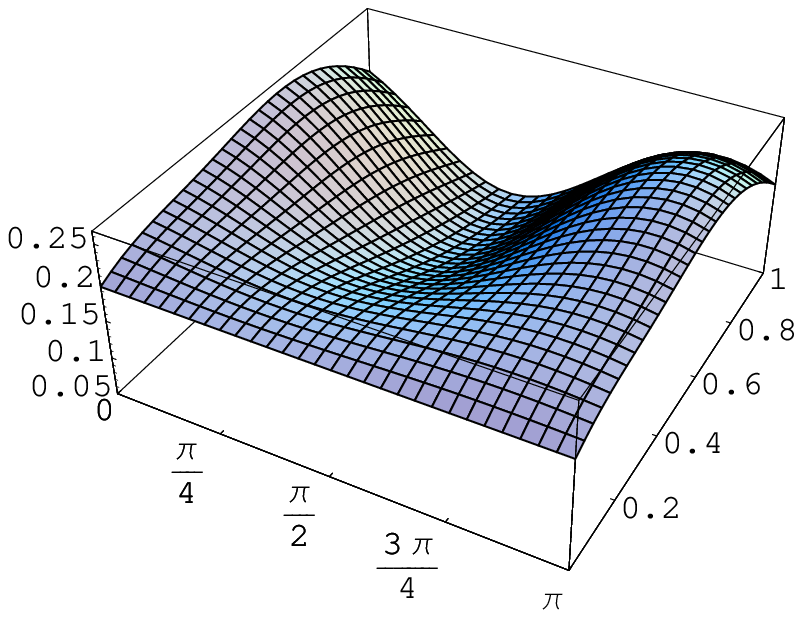}~\quad
   \includegraphics[width=15pc,height=10pc]{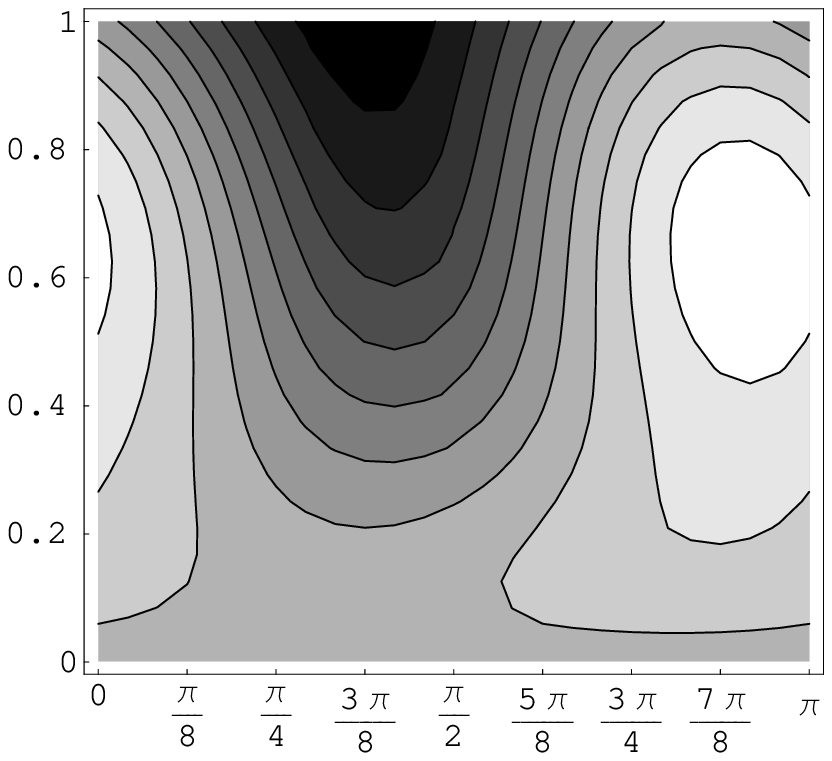}
\put(-385,68){$\mathcal{F}_\theta$}
    \put(-350,130){$(c)$}
    \put(-110,130){$(d)$}
        \put(-320,15){$\theta$}
     \put(-210,40){$\delta$}
      \put(-90,-9){$\theta$}
     \caption{The same as Fig.(1) but  for fixed $\Omega'_0=0.5$, and   $\phi=\pi/4,\pi/2$ for Figs.2(a,b), Figs.2(c,d), respectively.}
       \end{center}
\end{figure}

The effect of the larger value of the pulse strength
$(\Omega'_0=0.9$) is displayed in Figs.(1c$\&1d$). It is clear
that by increasing $\Omega'_0$, the Fisher information increases
even within larger values of the detuning.  Also,
$\mathcal{F}_\theta$ decays gradually  for $\theta\in[0,\pi/2]$
but its upper bounds are larger than those predicted at
$\Omega'_0=0.3$.  The Fisher information  increases as the initial
weight parameter increases, namely  for $\theta \in[\pi/2,~\pi]$.
Figs.(1c$\&1d$) display explicitly the upper and lower bounds of
the Fisher information. The most bright regions indicate that one
can estimate the weight parameter $\theta$ with high precision.

From Fig.(1) one may conclude the following: one may estimate the
weight parameter $(\theta)$ with hight degree of precision at
small values of the detuning parameter. This precision decreases
as one increases the value of the detuning parameter. However, one
may improve the degree of estimation by increasing the pulse
strength. If the initial qubit system encodes only classical
information (i.e. $\theta=0$ or $\pi$), the possibility of
estimating the weight parameter $(\theta)$ is larger than that
predicted if the initial qubit encodes quantum information (i.e
$0<\theta<\pi)$.

 Fig.(2),  shows  the effect of different values of the
phase parameter  $\phi=\pi/4,\pi/2$ on the precision of estimating
the weight parameter  $(\theta)$.  The general behavior is similar
to that displayed in Fig.(1c,d), but with  smaller values. It is
seen that the upper bounds of $\mathcal{F}_\theta$ at $\phi=\pi/4$
are larger than that for $\phi=\pi/2$.

From Figs.(1$\&2$), one observes that the possibility of
increasing the estimating rate of the weight parameter $(\theta)$
depends on the structure of the initial state, small values of the
detuning parameter and larger values of the pulse strength.

\item{\it Estimating the phase parameter ($\phi)$:} \\ Fig.(3)
displays the behavior of Fisher information $(\mathcal{F}_\phi)$
with respect to the phase parameter $\phi$  against the detuning
parameter $\delta$  for fixed values of the pulse strength,
$\Omega'_0$. It is assumed that the qubit system is initially
prepared in the state
$\psi(0)=\frac{1}{\sqrt{2}}(\ket{0}+e^{i\phi}\ket{1})$, namely,
the weight parameter $\theta=\pi/4$. The behavior of
$\mathcal{F}\phi$ shows that for small values of the detuning
parameter, Fisher information increases as the phase parameter
increases to reach it maximum values at $\phi=\pi/2$ and decreases
gradually to vanish completely at $\phi=\pi$. This behavior is
repeated in the interval of $\phi\in[\pi,2\pi]$, where the maximum
value is reached at $\phi=3\pi/2$ as shown in Fig.(3a). However,
as $\delta$ increases Fisher information decreases gradually and
completely vanishes at $\delta>0.6$.  From Fig.(3b) it is clear
that the upper bounds of $\mathcal{F}_\phi$ are shifted as
$\delta$ increases.

In Fig.3(c,d), we increase the pulse strength, where we set
$\Omega'_o=0.6$. The general behavior of the Fisher information
 is similar to that depicted  in Fig.3(a,b),
but the upper bounds of $(\mathcal{F}_\phi)$ is larger. On the
other hand, larger pulse strength protects the vanishing of the
Fisher information  with  larger values of the detuning parameter.

\begin{figure}[t!]
\begin{center}
 \includegraphics[width=15pc,height=12pc]{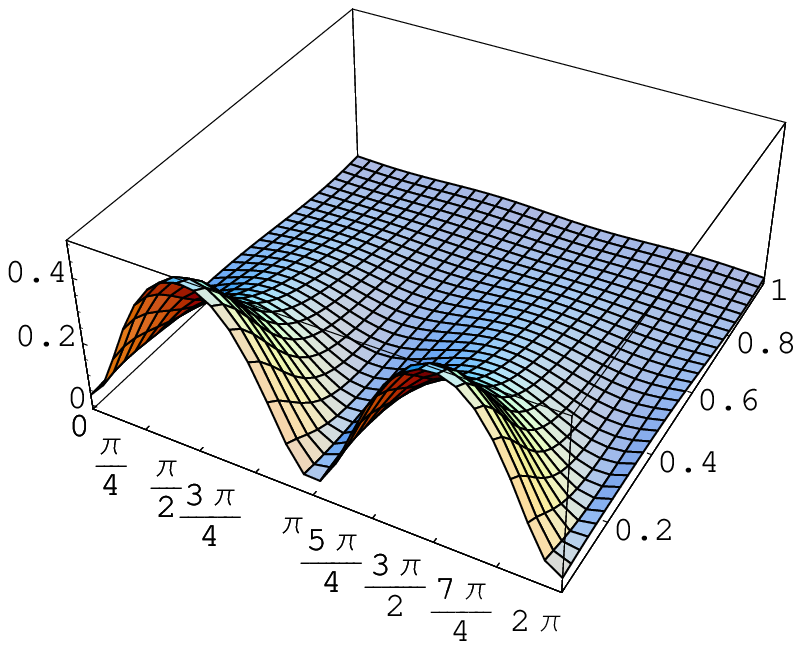}~\quad
   \includegraphics[width=14pc,height=10pc]{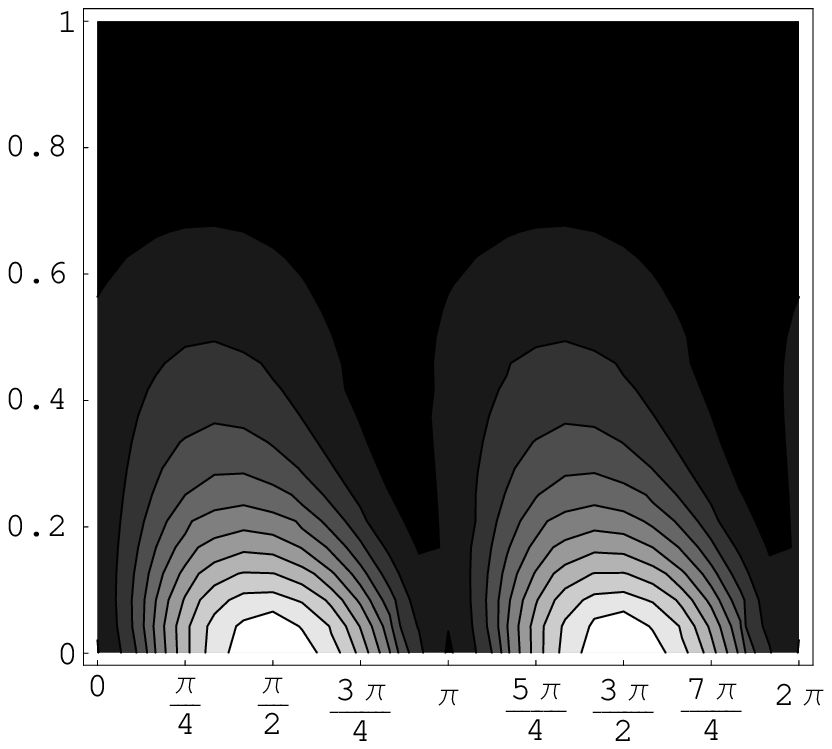}
    \put(-380,68){$\mathcal{F}_\phi$}
    \put(-350,130){$(a)$}
    \put(-110,130){$(b)$}
        \put(-320,15){$\phi$}
     \put(-195,43){$\delta$}
      \put(-80,-8){$\phi$}
        \put(-165,60){$\delta$}\\
   \includegraphics[width=15pc,height=12pc]{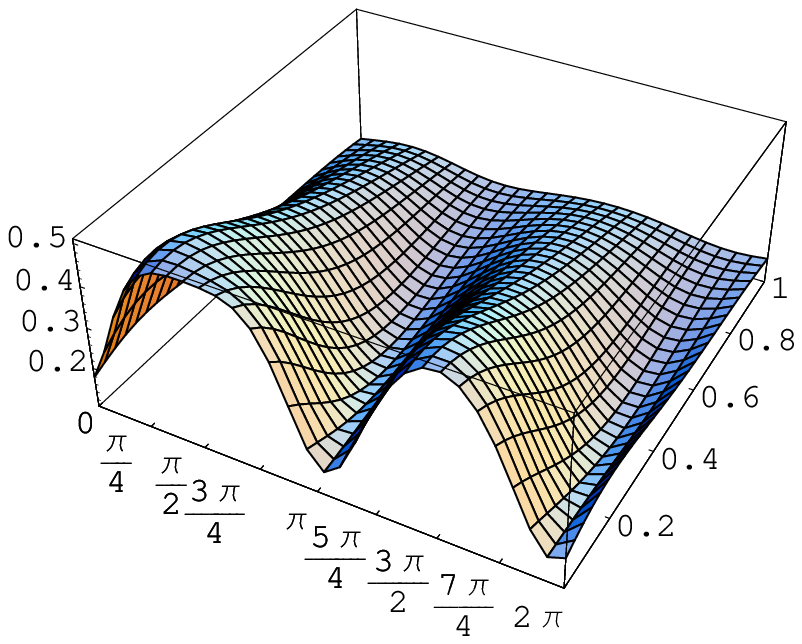}~\quad
   \includegraphics[width=14pc,height=10pc]{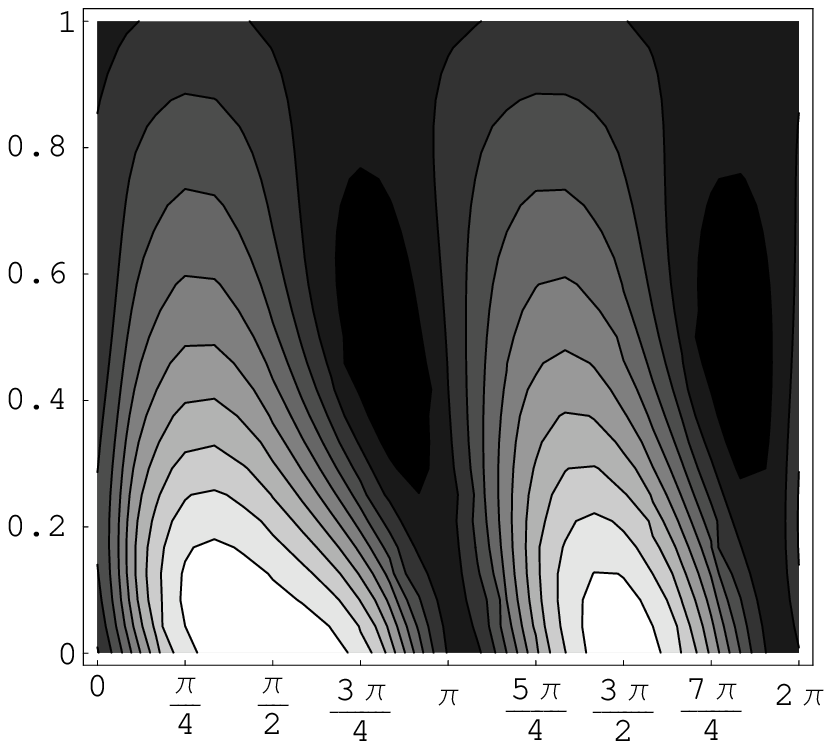}
   \put(-350,130){$(c)$}
    \put(-110,130){$(d)$}
    \put(-320,10){$\phi$}
     \put(-195,43){$\delta$}
      \put(-80,-8){$\phi$}
        \put(-165,60){$\delta$}
   \put(-380,68){$\mathcal{F}_\phi$}\\
 \includegraphics[width=15pc,height=12pc]{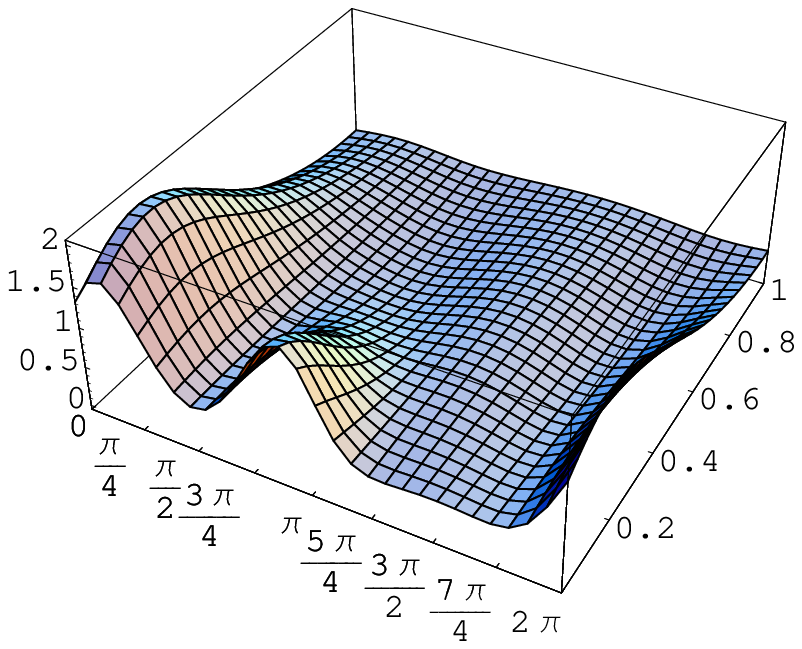}~\quad
   \includegraphics[width=14pc,height=10pc]{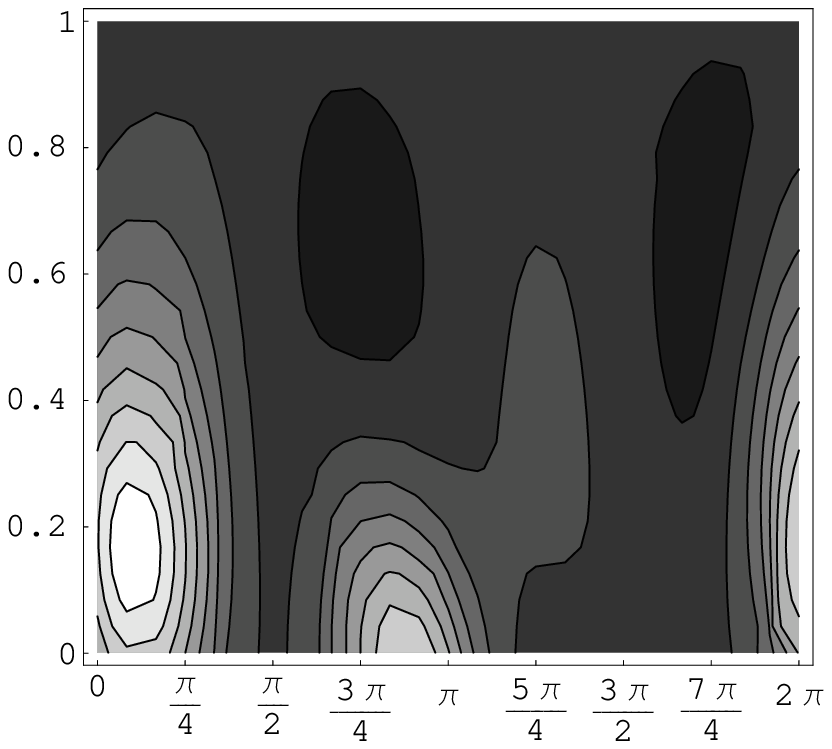}
   \put(-350,130){$(e)$}
    \put(-110,130){$(f)$}
    \put(-320,10){$\phi$}
     \put(-195,43){$\delta$}
      \put(-80,-8){$\phi$}
        \put(-165,60){$\delta$}
   \put(-380,68){$\mathcal{F}_\phi$}\\

 \   \caption{Fisher information $(\mathcal{F}_{\phi})$ in the rectangular pulse case with respect to the parameter $\phi$ and  the
   detuning parameter $\delta$ for fixed  $\theta=\pi/4$,  and $\Omega'_0=0.3,0.6, 0.9$ for the figures (a,b),
   (c,d),  (e,f), respectively.}
   \end{center}
\end{figure}

\begin{figure}[t!]
\begin{center}
 \includegraphics[width=15pc,height=12pc]{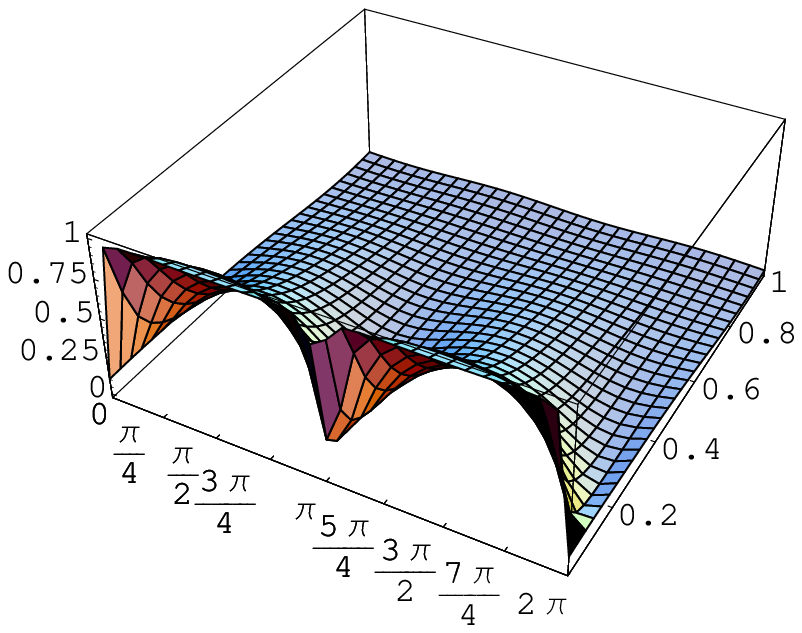}~\quad
   \includegraphics[width=14pc,height=10pc]{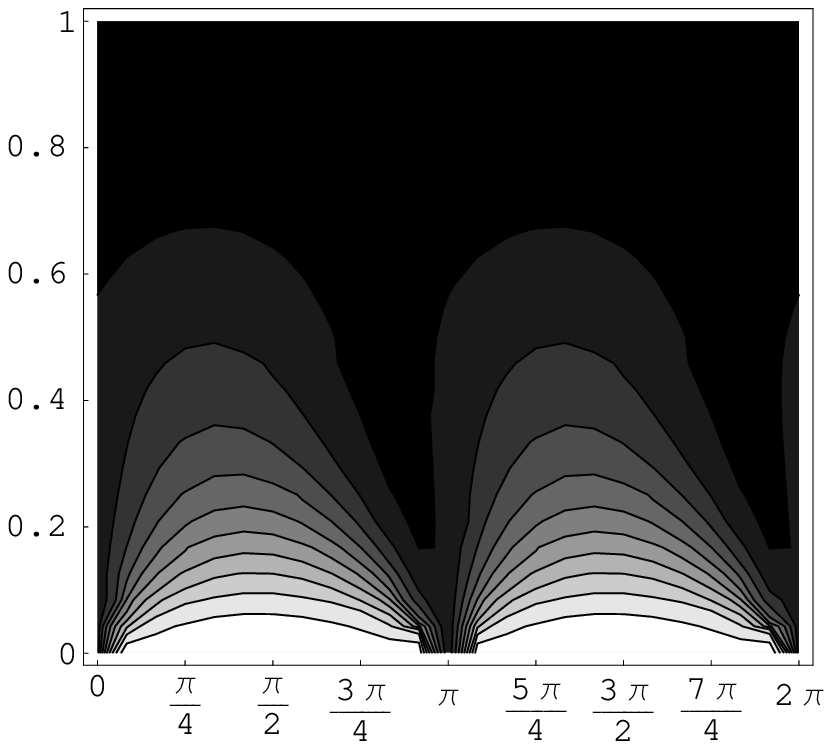}
    \put(-380,68){$\mathcal{F}_\phi$}
    \put(-350,130){$(a)$}
    \put(-110,130){$(b)$}
        \put(-320,15){$\phi$}
     \put(-200,43){$\delta$}
      \put(-80,-8){$\phi$}
        \put(-165,60){$\delta$}\\
 \includegraphics[width=15pc,height=12pc]{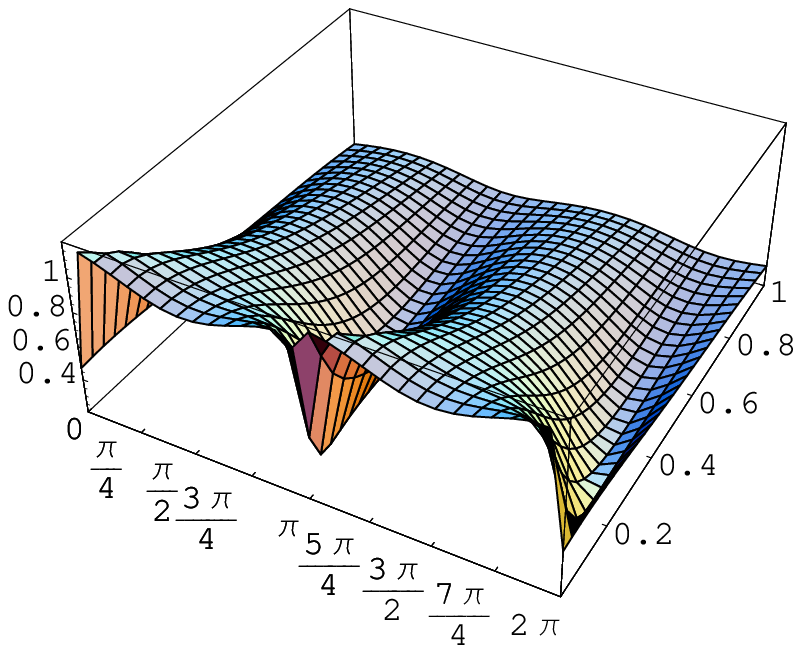}~\quad
   \includegraphics[width=14pc,height=10pc]{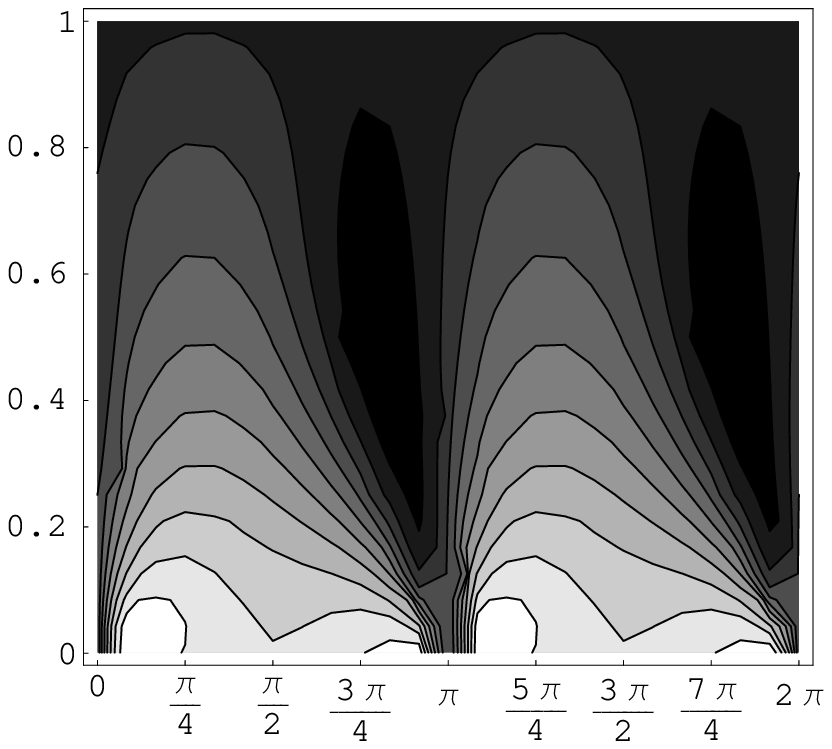}
    \put(-380,68){$\mathcal{F}_\phi$}
    \put(-350,130){$(a)$}
    \put(-110,130){$(b)$}
        \put(-320,15){$\phi$}
     \put(-200,43){$\delta$}
      \put(-80,-8){$\phi$}
        \put(-165,60){$\delta$}\\
        \includegraphics[width=15pc,height=12pc]{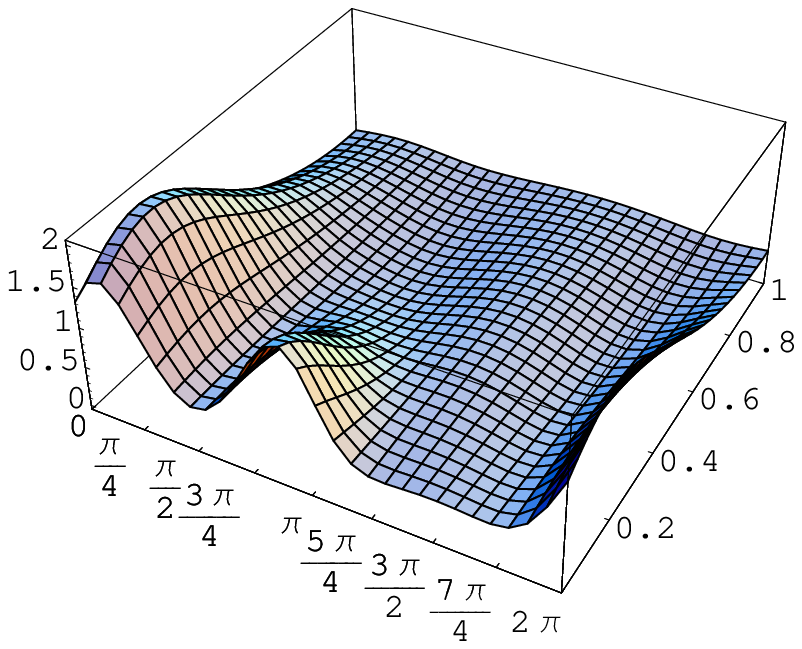}~\quad
   \includegraphics[width=14pc,height=10pc]{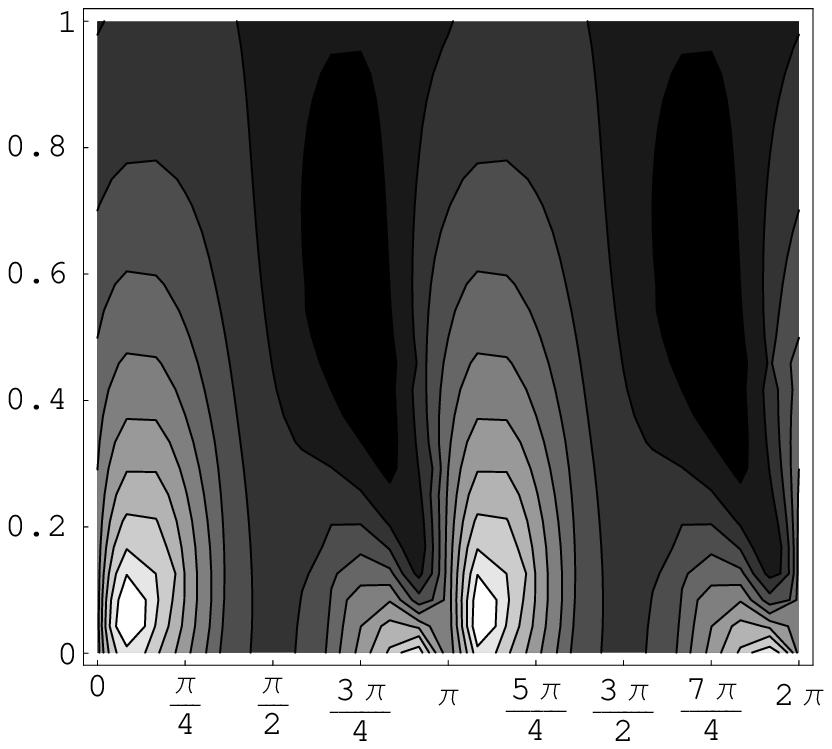}
    \put(-380,68){$\mathcal{F}_\phi$}
    \put(-350,130){$(c)$}
    \put(-110,130){$(d)$}
        \put(-320,15){$\phi$}
     \put(-200,43){$\delta$}
      \put(-80,-8){$\phi$}
        \put(-165,60){$\delta$}

     \caption{ The same as Fig.(3) but  for   $\theta=\pi/2$. }
   \end{center}
\end{figure}

In Fig.(4) we investigate the behavior of the Fisher information
$(\mathcal{F}_\phi)$ with respect to the phase parameter for a
system is initially prepared in the state
$\ket{\psi_0}=ie^{i\phi}\ket{1}$, namely, $\theta=\pi/2$. This
behavior shows that as soon as the system is pulsed at zero
detuning, the Fisher information increases suddenly to reach its
maximum values $(\mathcal{F}_\phi=1)$. As  $\phi$ increases the
upper values of $\mathcal{F}_\phi$ slightly decrease. However as
$\phi\rightarrow \pi$, the Fisher information decreases suddenly
to reach its minimum value at $\phi=\pi$. Similar to the behavior
in Fig.(4), $\mathcal{F}_\phi$ decreases as the detuning $\delta$
increases, but the upper bounds are larger than that displayed in
Fig.3.

From Figs.(3,4), one may conclude that the  possibility of
estimating the phase parameter $(\phi)$ decreases as  detuning
parameter increases. This possibility can be improved by
increasing the pulse strength. By increasing the weight parameter
$(\theta)$ the possibility of estimating the phase parameter
$(\phi)$ increases.

\end{enumerate}
\subsection{Exponential  and $sin^2$-pulses}
With   the initial qubit system is prepared in the state (15),
where the initial Bloch vector in the coherent state
$\ket{\theta,\phi}$, i.e.
$\row{s(0)}=(\cos\phi\sin\theta,\sin\phi\sin\theta,-\cos\theta)$,
and hence  $|\row{s(0)}|=1$. Using this initial value for
$\row{s(0)}$  in Eq.(11), one obtains the Bloch vector of the
final state in  a pure state, i.e. $|\row{s(t)}|=1$.  One can show
that $\mathcal{F}_{\theta,\phi}=1$ and hence it is independent of
the initial  parameters $\theta,\phi$. Therefore, in the case of
the Exponential and $\sin^2 $-pulses, one can estimate these
initial  parameters with high degree precision.

\section{conclusion}
The dynamics of a single qubit  initially prepared in a coherent
state $\ket{\theta,\phi}$ interacts with three different pulse
shapes, namely, rectangular, exponential and $sin^2$-pulses is
discussed, utilizing the  analytical solutions  in terms of the
corresponding  final Bloch vectors.

We investigate the  possibility of estimating the weight parameter
$(\theta)$ and the phase parameter $(\phi)$ of the initial single
qubit coherent state $\ket{\theta,\phi}$ under the effect of these
pulses. It is shown that the strength of the pulses have different
effect, while the initial phase angle $(\phi)$ has a similar
effect in  the three different cases as follows.

\begin{itemize}
\item For the rectangular pulse, the estimation degree of the
weight parameter $(\theta)$ decays as the atomic  deuning
parameter increases. This decay can be recovered by increasing the
value of the pulse strength. The upper bounds of estimation degree
decrease if the initial  qubit  prepared with a  suitable  phase
angle $(\phi)$, encodes quantum information.

\item The amount of Fisher information with respect to the phase
parameter  $(\phi)$ is  quantified for a system  which is
initially coded with quantum and classical  informations. It is
shown that, Fisher information $\mathcal{F}_\phi$ increases
suddenly at resonant case to reach its maximum value if the
initial phase $\phi=\pi/2$ and consequently one may estimate the
phase parameter with high degree of precision. However, as the
detuning increases the Fisher information decreases and therefore
the possibility of estimating the phase parameter decreases. If
the initial system is coded  with classical information, the upper
bounds of Fisher information for resonant and non-resonant cases
are  much larger.

\item  For  resonant  exponential  and $sin^2$ pulses, the Fisher
information is maximum and  consequently one can always estimate
the weight and the phase parameters $(\theta,\phi)$ with high
degree of precision.

\end{itemize}
{\it In conclusion}, the maximization of the estimation degree of
the weight parameter $(\theta)$  and the phase parameter $(\phi)$
of the single qubit depends on the
  structure of the initial state, small values of the detuning parameter and
larger values of the pulse strength. The pulse strength  is  the
effective control parameter to maximize the degree of estimation.


\begin{thebibliography}{nas}

\bibitem{Barnett} S. M. Barnett," Quantum Information (Oxford
Univ. Press, Oxford 2009).
\bibitem{Maccone} V. Giovannetti, S. Lloyd, and  L. Maccone, "
Quantum-Enhanced Measurements: Beating the Standard Quantum
Limit", Science {\bf 306},  1330-1336 (2004).


\bibitem{Jian011} J. Ma, Yi.-h Huang, X. Wang and C. P. Sun,"
Quantum Fisher informationof the Greenbereger-Horne-Zeillinger
state in decoherence channels", Phys. Rev. A {\bf 84} 022302
(2011).
\bibitem{Yao014} Q. Zheng, Y. Yao and Y. Li," Optimal quantum
channel estimation of two interacting qubits subject to
decoherence", Eur. Phys. J. D {\bf 68} 170 (2014).
\bibitem{Metwally017} N. Metwally,"Estimation of teleported and
gained parameters in a non-inertial frame", Laser Phys. Lett. {\bf
13}
 105206 (2016)


\bibitem{Wei013} W. Zhong, Z. Sun, J. Ma, X. Wang, and F. Nori,"
Fisher information under decoherence in Bloch representation",
Phys. Rev. A {\bf 87}, 022337 (2013).


\bibitem{Jing013}  J. Liu, X. Jing  and X. Wang ,"Phase-matching condition for enhancement of phase sensitivity in
quantum metrology", Phys. Rev. A {\bf 88}, 042316 (2013).

\bibitem{Wang011} H. Na Xiong and X. Wang",Dynamics of quantum
Fisher information in the Ising model", Physica A {\bf 390} 4719
(2011).


\bibitem{Hu014} G.-J. Hu and X.-X. Hu", Spin squeezing and quantum Fisher information for a mixed
Hamiltonian model", Int. J. Theor. Phys. {\bf 53} 533 (2014).


\bibitem{Ozaydin014} F. Ozaydin, A. A. Altintas, S. Bugu and C.
Yesilyurt," Behavior of quantum Fisher information of Bell pairs
under decoherence channels", Acta Phys. Pol A{\bf 125} 606 (2014).

\bibitem {Ali016} A. A. Altinatas," Quantum Fisher information of
an open and noisy syatem of the steady state", Annals of Physics
{\bf 367} 192 (2016).


\bibitem{yao014-1} Y. Yao, X. Xiao, Li Ge, X.G. Wang and C. pu
Sun,Quantum Fisher information in noninertial frames", Phys. Rev.
A. {\bf 89} 042336 (2014).

\bibitem{Metwally017-1}N. Metwally, "Unruh acceleration effect on the precision of parameter
estimation",    arXiv:1609.02092 (2016)


\bibitem{Metwally012}  N. Metwally and S. S. Hassan" Information
Transfer and orthogonality speed via-pulsed driven qubit",
Nonlinear Optics and Quantum Optics,{\bf 44} 267 (2012).




\bibitem{Metwally014}N. Metwally, H. A. Batarfi and S. S. Hassan,"
Long-lived entanglement with pulsed-driven initially entangled
qubit pair", Int. J. Quantum Infor. {\bf 12} 1450003 (2014).


\bibitem{Sukry008}  S. S. Hassan, A. Joshi and N. M. M.
Al-Madhari,"Spectrum of a pulsed driven qubit", J. Phys. B {\bf
41} 145503 (2008); and corrigendum: J. Phys. B {\bf 42} 089801
(2009).


\bibitem{Hassan010}S. S. Hassan, A. joshi and A. Batarfi" Spectrum
of triagnular pulsed-driven atom", Int. J. Therotical  Phys.,
Group Theory and Nonlinear Opt. {\bf 13} 371 (2010).


\bibitem{Bat012} H. A. Batarfi," Specturm of spin-$\frac{1}{2}$ system
driven by resonant exponential pulse" J. Nonlinear Opt. Phys. $\&$
Materials {\bf 21} 120025 (2012).


\bibitem{Bat017} H. A. Batarfi and S. S. Hassan" Haar wavelet
spectrum of an exponentially pulsed driven qubit", Nonlinear
Optics and Quantum Optics (2017), to appear.


\bibitem{al2015}R. A. Alharbey," Transient spectrum of $\sin^2-$
pulsed driven qubit",Int. J. Pure  $\&$App. Maths {\bf 110} 193
(2015).

\bibitem{al2016} R. A. Alharabey," Haar wevelet spectrum of
$sin^2-$ pulsed driven qubit", Optik {\bf 127} 9878 (2016).

\bibitem{Xing016} X. Xiao, Y. Yao, W.-J.Zhong, Y.-Ling and Y.-Mao
Xie," Enhancing teleportation of quantum Fisher information by
measurements", Phys. Rev. A {\bf 93}012307 (2016).



\end{thebibliography}
\end{document}